\documentstyle[12pt]{article}
\def\le{\langle}
\def\re{\rangle}

\def\N{\mbox{I\hspace{-.15em}N}}

\def\1{\mbox{I\hspace{-.15em}1}}

\def\Z{\mbox{Z\hspace{-.3em}Z}}

\def\R{{\rm I\hspace{-.15em}R}}

\def\b{\begin{equation}}
\def\e{\end{equation}}
\def\bee{\begin{enumerate}}
\def\eee{\end{enumerate}}
 
\title{Covariant two point function for \\  minimally coupled scalar
field \\in de Sitter space-time}
\author{ Mohammad Vahid Takook \thanks{e-mail: takook@ccr.jussieu.fr}}
\date{\today}
 
\begin{document}
\maketitle {\it 
\centerline{  Department of Physics, Razi University, Kermanshah, IRAN}
\centerline{ Laboratoire de Physique Th\'eorique de la Mati\`ere
Condens\'ee}
\centerline{ Universit\'e Paris $7$ Denis-Diderot,75251 Paris Cedex 05,
FRANCE}
}

\begin{abstract}

In a recent paper \cite{gareta1}, it has been shown that  negative norm states are indispensable for a fully covariant quantization of the minimally coupled scalar field in de  Sitter space. Their presence, while leaving unchanged the physical content of the theory, offers an automatic and covariant renormalization of the vacuum energy divergence. This paper is a completion of our previous work. An explicit construction of the covariant two-point function of the ``massless'' minimally coupled scalar field in de Sitter space is given, which is free of any infrared divergence. The associated Schwinger commutator function and retarded Green's function are calculated in a fully gauge invariant way, which also means coordinate independent. 
 
\end{abstract}
 
\vspace{0.5cm}
{\it Proposed PACS numbers}: 04.62.+v, 03.70+k, 11.10.Cd, 98.80.H
\vspace{0.5cm}

\section{Introduction}

The de Sitter metric appears as a logical geometrical background for inflationary cosmological scenario. We in particular insist on the fact that recent observational data are strongly in favor of a positive acceleration of the present universe. In a first approximation, this acceleration may be described by a constant $\Lambda$-term in the Einstein's equation of gravitation, even if this $\Lambda$-term may be permitted to slowly vary with time \cite{st,tu,boespost}. Therefore, the background space-time is very similar to a de Sitter space-time, and a quantization of the linear gravitational field without infrared divergence in such a space-time model may reveal to be of extreme importance for further developments. In a previous paper \cite{gareta1} we have shown that one can construct a covariant quantization of the
``massless'' minimally coupled scalar field in de Sitter space-time, which is free of any infrared divergence. In a forthcoming paper \cite{gareta2}, we prove that this is true for linear gravity (the traceless rank-2
``massless'' tensor field), which has been also demonstrated by \cite{hahetu, hiko}. Antoniadis, Iliopoulos and Tomaras \cite{anilto} have also  shown that the pathological large-distance behavior of the graviton propagator on a dS background does not manifest itself in the  quadratic part of the effective action  in the one-loop approximation. This means that the pathological behavior of the graviton propagator may be gauge dependent and so should not appear in an effective way as a physical quantity. Recently, de Vega and al. \cite{vera} have shown that,  in flat coordinates on de Sitter space-time, the infrared divergence does not appear in the ``massless'' minimally coupled scalar field. They use this result for studying the generation of the gravitational waves in de Sitter space-time. They have calculated the retarded Green's function in terms of flat coordinates of de Sitter space. 
  
This paper is a completion of our previous work \cite{gareta1}. We make explicit here the two-point function for the ``massless'' minimally coupled scalar field in de Sitter space-time in a fully gauge invariant way (coordinate independent), which is free of any infrared divergence. In Section $2$ we briefly recall the notations used in this paper.  In Section $3$, we give an explicit construction of the covariant two-point function and the related Schwinger commutator function. Then the retarded Green's function is calculated. In section $4$, we compare this result with the results obtained by de Vega and al., and we make precise in what way the latter are de Sitter invariant.  Finally, section $5$ contains a brief conclusion and outlook.
 
\section{Notation}
 
 The de Sitter space is an elementary solution of the cosmological
 Einstein equation. It is conveniently seen as a hyperboloid embedded in a
 five-dimensional Minkowski space
 \b X_H=\{x \in \R^5 ;x^2=\eta_{\alpha\beta} x^\alpha x^\beta
 =-H^{-2}\},\;\; \alpha,\beta=0,1,2,3,4, \e where
 $\eta_{\alpha\beta}=$diag$(1,-1,-1,-1,-1)$. The de Sitter metrics reads
 $$ds^2=\eta_{\alpha\beta}dx^{\alpha}dx^{\beta}=g_{\mu\nu}^{dS}dX^{\mu}dX^{\nu},\;\;\mu=0,1,2,3,$$ 
 where the $X^\mu$'s are the $4$ space-time coordinates in dS hyperboloid.
 Different coordinate systems can be chosen \cite{hael}. A so-called flat
 coordinatization satisfying $(1)$ and covering the half of the de Sitter hyperboloid is given by
         \b \left\{\begin{array}{clcr}  
  x^0&=H^{-1}\sinh Ht+\frac{1}{2}He^{Ht}\|\vec X\|^2 \\
 x^i&=e^{Ht}X^i\;\;,\;\;i=1,2,3 \\
 x^4&=H^{-1}\cosh Ht-\frac{1}{2}He^{Ht}\|\vec X\|^2.\\               
 \end{array} \right.\e
 Let us give details on this (non-global) coordinate system in order to compare our results with those of \cite{vera}. The line element is given by
 $$ ds^2=dt^2-e^{2Ht}(dX^1dX^1+dX^2dX^2+dX^3dX^3)$$ 
 \b =dt^2-e^ {2Ht }[d\rho^2+\rho^2( d \theta^2 + \sin^2 \theta
d\phi^2)],\e
 where $0\leq \rho<\infty$. In the conformal time notation, 
 $\eta=-H^{-1}e^{- Ht}\;,\;-\infty \leq \eta \leq 0$, and so we have
  \b ds^2=\frac{1}{H^2 \eta^2}  [d
          \eta^2- d \rho^2  - \rho^2(d \theta^2 + \sin^2 \theta
 d\phi^2)].\e 
 Given two points on the de Sitter hyperboloid $x$ and $x'$, the quadratic form
 \b {\cal Z}(x,x')=-H^2 x.x'=-H^2\eta_{\alpha \beta}x^\alpha x'^\beta,\e
is invariant under the isometry group $O(4,1)$. Moreover, any invariant function is function of this basic invariant ${\cal Z}(x,x')$. The latter can be written in terms of the invariant geodesic distance $\sigma$ between the points
 $x$ and $x'$.
 In the flat coordinate system $(2)$, we have
 $${\cal Z}(x,x')=-H^2 x.x'=1+\frac{H^2}{2} (x-x')^2 =1+\frac{1}{2\eta
 \eta'}[(\eta-\eta')^2-(\vec X-\vec X')^2 ]=\cosh H\sigma. $$
 We now briefly recall the quantization of the minimally coupled massless scalar field. The latter is defined by
 $$\Box_H \phi(x)=0,$$
 where $\Box_H$ is the Laplace-Beltrami operator on de Sitter space.
 As proved by Allen \cite{al}, the covariant canonical quantization procedure with positive norm states must fail in this case. The
 Allen's result can be reformulated in the following way: the Hilbert space generated by any complete set of modes is not de Sitter invariant, 
 $${\cal H}=\{\sum_k\alpha_k\phi_k;\;
 \sum_k|\alpha_k|^2<\infty\},$$
 where $k=\{(L,l,m)\in\N\times\N\times\Z;\; 0\leq l\leq L,\,
 -l\leq m\leq l\}$. This means that it is not closed under the action of the de~Sitter group. Nevertheless, one can obtain a fully covariant quantum field by adopting a new construction \cite{gareta1}. In order to obtain a fully covariant quantum field, we added all the modes $L< -1$. In consequence, we have to deal with an orthogonal sum of a positive definite inner product space and a negative one. This sum is closed under the de~Sitter group. The negative values of the inner product are precisely produced by the conjugate modes: $\le\phi_k^*,\phi_k^*\re=-1$. We do insist on the fact that the space of states contains unphysical states, which have a negative norm. Now, we can decompose the field into positive and negative norm parts (eq. $(30$) ref. \cite{gareta1})
   \b \phi(x)=\frac{1}{\sqrt{2}}\left[ \phi_p(x)-\phi_n(x)\right],\e
 where
 \b \phi_p(x)=\sum_{k }a_{k}\phi_{k}(x)+H.C.,\;\;
  \phi_n(x)=\sum_{k} b_{k}\phi^*(x)+H.C..\e
 The positive mode $\phi_p(x)$ is the scalar field as was used by Allen.
 Then we have the following operatorial relations
 \b    a_{k}|0>=0,\;\;[a_{k},a^{\dag} _{k'}]= \delta_{kk'},\;\;
   b_{k}|0>=0,\;\;[b_{k},b^{\dag} _{k'}]= -\delta_{kk'}.\e

 \section{Two-point function}
 
 The explicit knowledge of Wightman two-point function ${\cal W}$ allows us to make the quantum field formalism work. It also allows defining the various Green's functions of the wave equation. This function is defined by
 $$  {\cal W}(x,x')=<0\mid\phi(x)\phi(x')\mid 0>.$$
 For the scalar field $(6)$ we have
 $$ {\cal W}(x,x')=\frac{1}{2}[<0\mid\phi_p(x)\phi_p(x')\mid
 0>+<0\mid\phi_n(x)\phi_n(x')\mid 0>]
 $$
 \b=\frac{1}{2}[{\cal W}_p(x,x')+{\cal W}_n(x,x')], \e
 where ${\cal W}_n(x,x')=-{\cal W}_p^*(x,x')$ and ${\cal W}_p(x,x')$ is the two-point function for the positive modes as it was calculated by 
 Allen and Folacci \cite{alfo,fo}. Its expression reads
  \b {\cal W}_p(x,x')=\frac{H^2}{8\pi^2}[\frac{1}{1-{\cal Z}}-\ln (1-{\cal
 Z})+\ln  2+f_{AB}(\eta,\eta')], \e
 where $f_{AB}(\eta,\eta')$ is a function of the conformal time and is not
 de Sitter invariant. Therefore the two-point function $(9)$ is de Sitter invariant and reads:
 \b  {\cal W}(x,x')=\frac{iH^2}{8\pi} \epsilon (x^0-x'^0)[\delta(1-{\cal
 Z}(x,x'))-\theta ({\cal Z}(x,x')-1)], \e
 where 
  \b \epsilon (x^0-x'^0)=\left\{\begin{array}{clcr} 1&x^0>x'^0
 \\                                        
  0&x^0=x'^0\\  -1&x^0<x'^0.\\    \end{array} \right. ,\e
 and $\theta$ is the Heaviside step function. This expression in contrast to the previous calculation is de Sitter invariant, i.e. coordinate independent and also free of any infrared divergence. Now we can write the various
 Green functions from $(11)$. The anticommutation function is given by
    \b G^1(x,x')=\{ \phi(x),\phi(x') \}=2{\cal R}e {\cal W}(x,x')= 0. \e
 For the Schwinger commutator function, we have
 $$ G(x,x')=-i[ \phi(x),\phi(x') ]=- 2i{\cal W}(x,x')$$
 \b =\frac{H^2}{4\pi} \epsilon (x^0-x'^0)[\delta(1-{\cal Z}(x,x'))-\theta
 ({\cal Z}(x,x')-1)]. \e
 Then corresponding retarded Green's function is given by
 $$ G_R(x,x')=-\theta (x^0-x'^0) G(x,x')$$
 \b =\frac{-H^2}{4\pi}\theta (x^0-x'^0) \epsilon (x^0-x'^0)[\delta(1-{\cal
 Z}(x,x'))-\theta ({\cal Z}(x,x')-1)]. \e
 Thus, the retarded Green's function propagates on the past light cone and in its interior. It is also de Sitter invariant and free of any infrared divergence.
  
 \section{Comparison to previous works}
  
 The Schwinger commutator function $(14)$ is de Sitter invariant and so coordinate independent. In this section, we show that it is the same as the Schwinger commutator function which was calculated in the flat coordinate system $(2)$ by de Vega and al. \cite{vera}. The latter used the corresponding flat modes solutions. The field operator was written in terms of the Fourier transformation in the flat space 
 \b \phi(\eta, \vec X)=\int d^3\vec k [u_{\vec k}(\eta, \vec X)a(\vec
 k)+u_{\vec k}^*(\eta, \vec X)a^{\dag}(\vec k)],\e
 where
 \b u_{\vec k}(\eta, \vec
 X)=\frac{H}{(2\pi)^{3/2}\surd{2w}}(\eta-\frac{i}{w})\exp (-iw\eta+i\vec k
 \vec X),\e
 and $w= \mid \vec k \mid$. There results the Schwinger commutator
function
 $$ G(x,x')=-\frac{H^2}{4\pi \mid \vec X-\vec X' \mid} \times$$
 \b \left[ \eta
 \eta'(\delta(y_-)-\delta(y_+))+(\eta-\eta')(\theta(y_-)-\theta(y_+))+\frac{1}{2}(\mid
 y_+\mid - \mid y_- \mid)\right],\e
 where
 $$y_\pm=(\eta-\eta')\pm \mid \vec X-\vec X' \mid.$$
 For calculating this function the authors \cite{vera} have not used the two point function, which is divergence in their construction. They calculated directly the commutator function in which the infrared divergence disappears due to the sign of the divergence term. If we use the following relations,
  $$ \theta(t^2-x^2)=\theta(t-x)-\theta(t+x)+1,$$
 $$
 \epsilon(t)\theta(t^2-x^2)=\frac{1}{x}\{t[\theta(t-x)-\theta(t+x)]+\frac{1}{2}(\mid
 t+x  \mid-\mid t-x \mid)\},$$
  \b
 \epsilon(\eta-\eta')\theta(\frac{1}{2\eta\eta'}y_+y_-)=\frac{\eta-\eta'}{\mid
 \vec X-\vec X'  
  \mid}[\theta(y_-)-\theta(y_+)]+\frac{\mid y_+\mid - \mid y_- \mid
 }{2\mid \vec X-\vec X'  \mid},\e
  \b \delta(1-{\cal
 Z})=\delta(\frac{1}{2\eta\eta'}y_+y_-)=2\eta\eta'\delta(y_+y_-)=\frac{\eta\eta'}{\mid
 \vec X-\vec X' \mid}(\delta(y_-)+\delta(y_+)),\e
 it is easy to show that  Eq. $(18)$ is the same as Eq. $(14)$, which was calculated from the finite two point function $(11)$. The result of de Vega and al., which was claimed by them as free of infrared divergence, can be considered as a direct consequence of a requirement of de Sitter fully invariance. Note that the latter is irremediably broken if commutations are worked out with 
\underline{non-global} coordinates.

 \section{Conclusion}

 In this paper we have presented a covariant two-point function for minimally coupled scalar field in de Sitter space-time, which is free of any infrared divergence and is also de Sitter invariant (coordinate independent). Using this result in a forthcoming paper \cite{gareta2}, we are able to get a covariant two-point function for linear gravity (the traceless rank-2 ``massless'' tensor field) in de Sitter space-time, which is free of infrared divergence.
 This means that pathological behavior may be gauge dependent. It should not appear in an effective way in a physical quantity whereas it exists in an irreducible way in the pure-trace part (conformal sector) \cite{gareta2}. So far the physical meanings of these results have not been clarified. There are different possibilities. Let us mention two of them. The negative norm states are totally part of the structure of the renormalized field and it is not needed to remove them. Then we have de Sitter invariance for linear gravity (the traceless rank-2 ``massless'' tensor field). But in the inflationary model one introduces an inflaton scalar field. Because of this field, the conformal sector (pure trace) of the metric becomes dynamical and it must be quantized \cite{anmo2, anmamo}. This sector produces a gravitational instability and the breaking of de Sitter invariance. In this point of view, one may use a scalar-tensor theory for gravity $(g_{\mu \nu}, \phi)$ and this paves the way to the existence of an universe with a positive acceleration \cite{boespost}. The other week possibility is the following. Like for covariant   QED {\it \`a la Gupta-Bleuler} in Minkowski space negative norm states appear in a covariant quantization of the minimally coupled scalar field in de Sitter space. They are eliminated under the constraint of primordial fluctuations in the inflationary model. The latter indeed selects the physical states (the positive norm states) \cite{gaporeta}. The price to pay is precisely the breaking of de Sitter invariance. 
One can say that the negative norm states play the role ghost states. Now, breaking the de Sitter invariance is responsible of infrared divergence and this produces a gravitational instability. This gravitational instability and the primordial quantum fluctuations of the inflaton scalar field define the inflationary model. The latter can explain the formation of the galaxies, clusters of galaxies and the large scale structure of the universe \cite{lepost}.
 
 \vskip 0.5 cm
 
 \noindent {\bf{Acknowlegements}}: The author would like to thank J. P.
 Gazeau, J. Iliopoulos, D. Polarski and J. Renaud for very useful
discussions.

 \end{document}